\documentclass[pra,showpacs]{revtex4}
\usepackage[english, russian]{babel}
\usepackage[cp1251]{inputenc}
\inputencoding{cp1251}
\usepackage{graphicx}
\usepackage{color}

\begin{document}



\def\BE{\begin{equation}}
\def\EE{\end{equation}}
\def\BY{\begin{eqnarray}}
\def\BEA{\begin{eqnarray}}\def\EY{\end{eqnarray}}\def\EEA{\end{eqnarray}}
\def\L{\label}
\def\nn{\nonumber}
\def\({\left (}
\def\){\right)}
\def\<{\langle}
\def\>{\rangle}
\def\[{\left [}
\def\]{\right]}
\def\o{\overline}
\def\BA{\begin{array}}
\def\EA{\end{array}}
\def\ds{\displaystyle}

\title{Purity and Covariance Matrix }

\author{ T. Golubeva, {Yu. Golubev}  }
\address{St.~Petersburg State University, \\
198504 Stary Petershof, St.~Petersburg, Russia}
\date{\today}

\begin{abstract}

Basing on the simplest single-mode field source, we investigate the role of the various covariance matrices for reconstructing the field state and
describing its quantum statistical properties. In spite of the fact that the intracavity field is a single-mode field, we take into account the
natural multimode structure arising in the field, when it leaves the cavity for the free-space propagation. We show how the purity of the field state
can be calculated using the different covariance matrices.

\end{abstract}


\maketitle

\medskip


\section{Introduction}

In order to use the advantages inherent in the non-classical world, to improve the measurement technique with high resolution or quantum information
protocols, it is often necessary to complete knowledge of the quantum state of the electromagnetic field. Methods to characterize fully the quantum
state were part of a set of tools of quantum optics for two last decades \cite{1,2}.

In quantum optics as well as in any theory of random processes, so-called covariance matrix (CM) is widely used. The matrix is constructed on the
basis of pair correlation functions, and, in the case of fields with Gaussian statistics, it gives a comprehensive description of these fields.
Description of the statistical properties of the field from some specific source is one of the possible applications of the CM. The second is a
theoretical reconstruction of the quantum state with the help of experimentally measured CM.

In many applications the use of the CM is convenient and productive. However, when we have to take into account the multi-mode structure of the field
we face some difficulties with CM interpretation. In this situation, one can carry out several different measurements under the system, and each of
them implies its own CM. Thus, there is a possibility to construct not only one unique CM, but several ones. The questions naturally arising are
'Which of these matrices contain a comprehensive information about the field? Should we use for the reconstruction of the state all of them or only
some one?' In this article we will try to answer these questions on the simple example of a single-mode resonator source.

Just as in recent paper \cite{3}, we come to the conclusion that the description within the limit of the balanced homodyne detection in terms of the
spectral transformation of amplitude quadrature components is insufficient, because this is unsuitable for monitoring one selected mode. Using the
preliminary spectral decomposition of the field and selecting only one spectral component on the detector, one can overcome this difficulty. This
procedure can be realized, for example, in the scheme of the resonator detection \cite{4,5}.

Another issue that is usually interested in quantum optics, is a purity of the quantum state. Due to the fact that, in the case of the Gaussian
statistics, the CM ensures the complete description of the field, we can expect that the purity of the state may be obtained directly from the CM.
There is a well-known formula by which the state purity is inversely proportional to the square root of the CM determinant. However, for the
multimode field, this question is not obvious, at least because one can produce not an unique CM but several different ones. We will discuss this
problem in Sec. \ref{Last}.

\section{Single-mode intracavity field \L{II}}

Here we want to discuss once more the approach, widely exploited in experimental and theoretical physics and associated with the covariance matrix.
This matrix is constructed on the basis of pair correlation functions, and thus contains, in general, limited information on the statistical picture
of the field. However, for fields with Gaussian statistics, it turns out to be exhaustive. Thereby under investigation of the Gaussian fields, a
description by CM is very convenient, because, on the one hand, this can be observed experimentally, and, on the other hand, provides a complete
information. However, as we shall see below, the actual situation is not so simple, if one takes into account the multimode structure of the field.
The fact that almost always we need to consider the fields available for the measurement as multimode. As a result, there are many variants of the
measurement procedures as for any system with many degrees of freedom. For each of them its own CM may be recorded but we can not say that any one of
them provides a complete description of the statistical properties of the field.

To illustrate this, let's start with considering a single mode electromagnetic field inside a high-Q cavity. We can interpret it as field oscillator
with its own frequency, which coincides with the frequency of the actual mode. Its state can be pure or mixed, depending on the specific conditions
within the cavity. It may have a classical or quantum properties. Let the steady state of the oscillator be described by the Wigner
quasi-distribution in the Gaussian form
\BY
&&W(x,y)=\frac{2}{\pi}\sqrt{ab-c^2}\;\exp\( -[a\;\delta x^2+b\;\delta y^2 -2c\;\delta x\delta y]\).\L{1}
\EY
Here $a,b>1,\;ab>c^2$. For the oscillator in a vacuum state  $a=b=1, \; c=0$.  The quadratures $x$ and  $y$ are introduced as real and imaginary
parts of this value
\BY
&&\delta\alpha=\frac{1}{\sqrt2}(\delta x+i\delta y)\L{12},\qquad\alpha=\o\alpha+\delta\alpha.\L{2}
\EY
Here $\alpha$ is eigen-number of the non-hermitian annihilation operator $\hat a |\alpha\rangle=\alpha|\alpha\rangle$. It is not difficult to see
that the pair momenta of the introduced quasi-distribution is given by
\BY
&&2\;\o{\delta x^2}={b}\det{\cal M},\qquad 2\;\o{\delta y^2}={a}\det{\cal M} ,\qquad2\;\o{\delta x\delta y}=
{c}\det{\cal M},\quad \mbox{where}\quad\det{\cal M}=\frac{1}{ab-c^2}.\L{3}
\EY
For example, if one obtains experimentally the CM
\BY
{\cal M}=\left(%
\begin{array}{cc}
  2\;\o {\delta x^2} & 2\;\o{ \delta xy}\\
 2\;\o{ \delta xy} &2\;\o{ \delta y^2} \\
  \end{array}
\right),\L{3}
\EY
then the parameters $ a, b $ and $ c $ can be expressed via the matrix elements of this CM in the form
\BY
&&a={2\;\o{\delta y^2}}/{\det{\cal M}},\quad b=2\;\o{\delta x^2}/\det{\cal M},\quad c=2\;\o{\delta x\delta
y}/\det{\cal M},
\EY
where
\BY
&&\det{\cal M}=4\(\o {\delta x^2}\;\o {\delta y^2}-\o {\delta x\delta y}^2\).
\EY
Thus, one can reconstruct the Wigner function (\ref{1}).

It is clear that the CM, in principle, makes it possible to describe all the statistical properties of the oscillator, and in this sense it is no
worse than the Wigner function itself. For example, the purity of the quantum state of the oscillator may be derived via the CM
\BY
&&{\cal P}=\frac{1}{\sqrt{\det {\cal M}}}.\L{5}
\EY
At the same time, we know that the single-mode state purity    is expressed via the Wigner function as
\BY
&&{\cal P}=\pi\int d^2\alpha\; W^2(\alpha).\L{}
\EY
Substituting here the Wigner function (\ref{1}) one can obtain after the direct integration ${\cal P}=\sqrt{ab-c^2}$, and formula (\ref{5}) is
confirmed.

Thus we can conclude that if we could measure the intra-cavity field, then following experimentally the elements of covariance matrix (\ref{3}), we,
in fact, would receive a full description of the (Gaussian) statistical properties of the quantum state and estimate its purity by using Eq.
(\ref{5}).

As a result of the other experiment, we can get another CM constructed directly on the complex amplitudes $\alpha$ (and not on their quadratures as
before)
\BY
{\cal M}=\left(%
\begin{array}{cc}
  \o {2|\delta \alpha|^2} & 2\o{ \delta\alpha^2}\\
 2\o{ \delta\alpha^{\ast2}} &2\o{| \delta\alpha|^{2}} \\
  \end{array}
\right).\L{8}
\EY
Taking into account equality (\ref{2}), the Wigner function (\ref{1}) in the new variables is given by
\BY
&&W(\alpha)=\frac{2}{\pi}\sqrt{ab-c^2}\;\exp\[-(a+b)\;|\delta \alpha|^2-\[(a-b)/2+ic\]\;\delta \alpha^2 -
\[(a-b)/2-ic\]\;\delta \alpha^{\ast2}\].\L{9}
\EY
The CM (\ref{8}) allows us to reconstruct quasi-distribution (\ref{9}) completely, because we have the equalities
\BY
&& a+b= 4{\o{|\delta \alpha|^2}}/{\det{\cal M}},\qquad (a-b)/2+ic=-2{\o{\delta \alpha^{\ast2}}}/{\det{\cal M}},\qquad
(a-b)/2-ic=-2{\o{\delta \alpha^2}}/{\det{\cal M}},
\EY
where
\BY
&& \det{\cal M}=4\(\o{|\delta \alpha|^2}^2-|\o{\delta \alpha^2}|^2\),
\EY
With the same success CM (\ref{8}) allows the reconstruction of the quasi-distribution (\ref{1}). The same is true for the matrix (\ref{3}), it also
makes it possible to reconstruct both distributions.

Concluding this section, we can say that in the case of a single oscillator we can reconstruct the Wigner function for this oscillator using any
representation of the covariance matrix.

\section{Multimode structure of a field}

In the real experiment, the measuring procedure is carried out not with the intracavity field, but with the field that came out in free space. When
field crosses the output mirror, it acquires some new characteristics compared with intracavity situations. First, the statistical properties of the
light can be significantly changed  due to vacuum fluctuations. Second, the mode structure of the field turns out to be different, as there are a lot
of the free space modes (continuum) on the spectral width of the cavity mode. Thus even if inside the cavity only one mode is excited, outside the
field is obviously multimode.

Take into account this multimode structure, we can derive the field in the form of the quasi-monochromatic plane wave:
\BY
&&\hat E(z,t)=i\sqrt{\frac{\hbar\omega_0}{2\epsilon_0cS}}\;e^{ik_0z-i\omega_0t}\hat A(t).\L{11}
\EY
Here $k_0,\;\omega_0$ are z-component of the wave vector and the corresponding mode frequency, $ c $ is the light velocity  in vacuum, $ S $ is an
effective cross section of the light beam. The slowly varying in time Heisenberg amplitude $ \hat A (t) $ obeys the canonical commutation relations
\BY
&&\[\hat A(t),\hat A^\dag(t^\prime)\]=\delta(t-t^\prime),\qquad
\[\hat A(t),\hat A(t^\prime)\]=0.\L{12}
\EY
The amplitude $ \hat A $ is normalized so that the value  $ \langle \hat A^\dag \hat A \rangle $ has a sense of the
mean photon flux, i.e., the average number of photons running through the cross-section of the beam $ S $ per {\it
sec}.

Let us rewrite commutation relation (\ref{12}) for the Fourier components of the field amplitude
\BY
&&\hat A_\Omega=\frac{1}{\sqrt{2\pi}}\int dt\hat A(t)e^{\ds i\Omega t},\mbox{where}\quad \hat
A(t)=\frac{1}{\sqrt{2\pi}}\int d\Omega\hat A_\Omega e^{\ds -i\Omega t}. \L{}
\EY
Quantizing  frequency scale, we obtain
\BY
&&\[\hat A_\Omega,\hat A^\dag_{\Omega^\prime} \]=\delta_{\Omega\Omega^\prime}.\L{}
\EY

Hereinafter we assume that the state of field outside the cavity obeys the Gaussian statistics. The appearance of
many degrees of freedom in the system expands the class of possible measurement procedures, for each of which its own
CM can be mapped.

Let's consider the possibility of reconstruction of the state function (Wigner quasi-distribution), referring to the results of some mental
experiments. Because our system is composed of a set of field oscillators, then, in principle, we should talk about the reconstruction of states of
different subsystems consist of one, two, three, etc. oscillators.

We consider two approaches to the study of the multimode field, based on the balanced homodyning. In the first approach, the field leaving the cavity
is previously resolved into spectral decomposition, and only one spectral component is photodetected. In this case, the observed values are the
imaginary and the real parts of the spectral amplitude:
\BY
&&\hat A_\Omega=\frac{1}{\sqrt2}(\hat Q_\Omega+i\hat P_\Omega),\quad\mbox{where}\quad \hat Q_\Omega^\dag=\hat
Q_\Omega,\qquad\hat P_\Omega^\dag=\hat P_\Omega,\qquad\[\hat Q_\Omega,\hat P_{\Omega}\]=i .\L{15}
\EY
This approach can be practically realized in the technique of the resonator detection \cite{4,5}.

In the second approach, the entire light from the cavity falls on the photodetector, and, in this case, we follow the
other values, namely the quadrature component of the field, which are introduced as the real and imaginary parts of
the field amplitude
\BY
&&\hat A(t) = \frac{1}{\sqrt2} \(\hat X(t)+ i \hat Y(t)\),\quad\hat X(t)=\hat X^\dag(t),\quad\hat Y(t)=\hat
Y^\dag(t).\L{16}
\EY
Further we are following not quadratures $ \hat X (t) $ and $ \hat Y (t) $ themselves, but their Fourier transforms $ \hat X_\Omega $ and $ \hat
Y_\Omega $:
\BY
&&\hat A_\Omega = \frac{1}{\sqrt2} \(\hat X_\Omega+ i \hat Y_\Omega\),\quad\hat X_\Omega=\hat X_{-\Omega}^\dag,\quad\hat Y_\Omega=\hat
Y_{-\Omega}^\dag.\L{19}
\EY
This spectral components are non-Hermitian in distinguish from operators $ \hat P_\Omega $ and $ \hat Q_\Omega $.

\section{Following $\hat P_\Omega$ и $\hat Q_\Omega$}
\subsection{Reconstruction of the single-mode state}

Since the investigated field is multimode,  we can supply different questions in dependent on our specific interest. For example, we may be
interested in the quantum state of the selected field oscillator or a combination of two or more oscillators. If the light in the scheme of balanced
homodyning is detected completely, the photocurrent operator reads
\BY
&&\hat i(t)=A_{LO}^\ast \hat A(t)+ A_{LO} \hat A^\dag(t),\L{20}
\EY
where $ A_{LO} $ is the local oscillator amplitude.

One can see, in this case, all the spectral components of the field are involved in the formation of the
photocurrent.

At the same time, one can consider another version of the experiment, when after a preliminary spectral decomposition of the field, only one spectral
component of the field allows for detecting. In this case, the photocurrent can be written as
\BY
&&\hat i(t)\sim A_{LO}^\ast \hat A_\Omega e^{-i\Omega t}+ A_{LO} \hat A_{\Omega}^\dag e^{i\Omega t}.\L{21}
\EY
It is clear that under this approach we are able to follow only one selected oscillator with frequency $ \omega_0 + \Omega $.

If a monochromatic modulation with frequency $\Omega$ of the local oscillator is applied, i.e., put $ A_{LO} \to A_{LO} e^{-i \Omega t} $, then the
photocurrent is given by
\BY
&&\hat i(t) \sim A_{LO}^\ast \hat A_\Omega + A_{LO} \hat A_{\Omega}^\dag .\L{22}
\EY
Now we have an opportunity to follow  the "quadratures" $\:$ of the  spectral amplitudes  that are Hermitian
operators and are introduced into consideration according to (\ref {15}) as the real and imaginary part of the value
$ \hat A_\Omega $.

Let us derive the local oscillator amplitude as
 \BY
&& A_{LO}=\frac{1}{\sqrt2}(  Q_{LO}+i  P_{LO}),\quad\mbox{where}\quad \hat Q_{LO}^\ast=\hat Q_{LO},\qquad\hat P_{LO}^\ast=\hat P_{LO} .
\EY
Then the photocurrent (\ref{20}) can be rewritten in the form
\BY
&& \hat i\sim Q_{LO}\hat   Q_\Omega+ P_{LO} \hat P_{\Omega}.
\EY
On this basis, one can carry out several measuring procedures with the proper local oscillator amplitudes and, as a
result, derive $2\times2$ CM:
\BY
{\cal M}_\Omega=\left(%
\begin{array}{cc}
  2\;\langle{\delta
\hat Q_\Omega^2} \rangle& 2\langle{\{\delta \hat Q_\Omega,\delta \hat P_\Omega}\}\rangle\\
  2\langle{\{\delta \hat Q_\Omega,\delta \hat P_\Omega}\}\rangle &2\; \langle{\delta
\hat P_\Omega^2}\rangle  \\
  \end{array}
\right).\L{22}
\EY
Let us reformulate this matrix in the Wigner domain
\BY
{\cal M}_\Omega=\left(%
\begin{array}{cc}
  2\;\langle{\delta
q_\Omega^2} \rangle& 2\langle{\delta q_\Omega\delta p_\Omega}\rangle\\
  2\langle{ \delta q_\Omega\delta p_\Omega}\rangle &2\; \langle{\delta
p_\Omega^2}\rangle  \\
  \end{array}
\right).\L{26}
\EY
Here the amplitudes $p_\Omega$ and $q_\Omega$ are introduced according to equalities
\BY
&&\hat A_\Omega|\alpha_\Omega\rangle=\alpha_\Omega|\alpha_\Omega\rangle,\qquad
\alpha_\Omega=\frac{1}{\sqrt2}\(q_\Omega+ip_\Omega\),\quad\mbox{where}\quad q_\Omega=q^\ast_\Omega,\quad p_\Omega
=p^\ast_\Omega. \L{27}
\EY
The formal scheme here is very similar to that discussed in Section \ref{II} for the intra-cavity mode field. To generalize the results obtained
there for this case, one should replace $ x\to q_\Omega$ and $y\to p_\Omega$. On this basis, we can conclude that the CM (\ref{26}) allows to
reconstruct the Gaussian distribution function of a single oscillator of multimode field, and the purity of the quantum state of this oscillator is
given by
\BY
&&{\cal P}_{\Omega}=\frac{1}{\sqrt{\det {\cal M}_{\Omega}}}.\L{23}
\EY

\subsection{Reconstruction of the two-mode states}

A study of the selected $(\omega_0+\Omega)$-oscillator is easily extended to the monitor of two arbitrary oscillators
with frequencies $\omega_0+\Omega_1 $ and $ \omega_0 + \Omega_2 $. For this case, one gets the $4\times4$ CM:
\BY
{\cal M}_{\Omega_1,\Omega_2}=\left(%
\begin{array}{cc}
  {\cal M}_{\Omega_1}& {\cal N}_{\Omega_1,\Omega_2}\\
  {\cal N}_{\Omega_1,\Omega_2}&{\cal M}_{\Omega_2}\\
  \end{array}
\right).\L{24}
\EY
Here $2\times2$ matrices $ {\cal M}_{\Omega_1} $ and $ {\cal M}_{\Omega_2} $ on the main diagonal are the same as the single-mode matrix (\ref {22})
for $ \Omega = \Omega_1 $ and $ \Omega = \Omega_2 $, respectively. The 2 by 2 matrix ${\cal N}_{\Omega_1,\Omega_2}$ reads
\BY
{\cal N}_{\Omega_1,\Omega_2}=\left(%
\begin{array}{cc}
  2\langle{\delta
q_{\Omega_1}\delta
q_{\Omega_2}} \rangle& 2\langle\delta q_{\Omega_1}\delta p_{\Omega_2}\rangle\\
  2\langle{\delta q_{\Omega_1}\delta p_{\Omega_2}}\rangle &2 \langle{\delta
p_{\Omega_1}\delta
p_{\Omega_2}}\rangle  \\
  \end{array}
\right)\L{25}
\EY
and describes all possible correlations between selected oscillators. For the stationary light flux
\BY
&&{\cal N}_{\Omega_1,\Omega_2}={\cal N}_{\Omega_1,-\Omega_1}\delta_{\Omega_1,-\Omega_2},
\EY
where $\delta_{\Omega,\Omega'}$ is a Kronecker symbol.

Therefore, when $ \Omega_1 \neq-\Omega_2 $,  the correlations are absent for the stationary flux and $ {\cal N}_{\Omega_1, \Omega_2} = 0 $. In
particular, this means that the purity of the system consisting of two independent oscillators is factorized in the form
\BY
&&{\cal P}_{\Omega_1,\Omega_2}={\cal P}_{\Omega_1}{\cal P}_{\Omega_2},
\EY
where ${\cal P}_{\Omega}$ is single-frequency purity according to Eq. (\ref{23}).

Thus for Gaussian statistics the purity of two arbitrary oscillators survives the same form as for the single
oscillator
\BY
&&{\cal P}_{\Omega_1\Omega_2}=\frac{1}{\sqrt{\det {\cal M}_{\Omega_1\Omega_2}}},\quad\mbox{where}\quad\det {\cal M}_{\Omega_1\Omega_2}=\det {\cal
M}_{\Omega_1}\det {\cal M}_{\Omega_2}, \L{28}
\EY
and equals to the product of two purities for single oscillators.

A special situation arises when we follow the two oscillators with the frequencies $ \Omega_1 $ and $ \Omega_2 = -
\Omega_1 $. Because under these conditions the correlation matrix (\ref{25}) is not equal to zero, we have no reason
to think that Eq. (\ref{28}) survives. A new expression could relatively easily be calculated. For that we need to
specify the appropriate two-mode Wigner quasi-distribution, provided the field statistics is Gaussian. However we
want to consider this task in the next section in the other measurement procedure.

Hence it follows that the covariance matrix constructed on the basis of real and imaginary parts of the spectral field amplitude, provide a
reconstruction of a Gaussian state of any subsystem of the observed field.

\section{Following  $\hat X_\Omega$ и $\hat Y_\Omega$ }\L{Last}

\subsection{Covariance matrix}

In the experimental technique within the balanced homodyning, the CM associated with the observation of the spectral components of the field
quadratures (\ref{19}) is often discussed. For example, in our work \cite{6} dedicated to the above-threshold non-degenerate parametric generation,
we have analyzed the CMs of this particular type. Further we will discuss what is the role of this matrix in the description of the field properties.
This is especially important as in the previous section we concluded that the parameters $\hat Q_\Omega$ and $\hat P_\Omega$ allow the reconstruction
of a Gaussian state of any field subsystem.

In this case, the photocurrent  is determined by the Hermitian operator (\ref{20}). Passing to Fourier domain, we obtain
\BY
&&\hat i_\Omega=A_{LO}^\ast \hat A_\Omega+ A_{LO} \hat A_{-\Omega}^\dag.\L{29}
\EY
With such a measurement procedure, two field oscillators with the frequencies $ \omega_0 \pm \Omega $ contribute to
each spectral component of the photocurrent $ \hat i_\Omega $, i.e., in this case, we follow the pairs of oscillators
with frequencies, disposed symmetrically with  respect to the mode frequency  $ \omega_0 $.

 Thus, this experiment does not make us a possibility to estimate the statistical
properties of the single oscillator, in contrast to the measurement that was discussed in the previous section.

The intrinsic variables in this approach are the Hermitian quadrature components of the field, that were introduced
above (\ref {16}) as the real and imaginary parts of the field amplitude.

Photocurrent spectrum (\ref {29}) can be expressed via the spectral components of the quadratures as
\BY
&&\hat i_\Omega= Q_{LO}\hat X_\Omega+P_{LO}\hat Y_\Omega,\L{30}
\EY
where the spectral quadrature $ \hat X_\Omega $ and $ \hat Y_\Omega $ are introduced in (\ref{19}). On this basis,
the $2\times2$ CM in the Wigner domain takes a form
\BY
{\cal M}_{\Omega,-\Omega}=\left(%
\begin{array}{cc}
  2\langle \delta x_\Omega \delta x_{-\Omega}\rangle & 2\langle \delta x_\Omega \delta y_{-\Omega}\rangle\\
  2\langle \delta x_\Omega \delta y_{-\Omega}\rangle  &2\langle \delta y_\Omega \delta y_{-\Omega}\rangle \\
  \end{array}
\right),\L{36}
\EY
where
\BY
x_\Omega=\frac{1}{\sqrt2}\( \alpha_\Omega+{\alpha}_{-\Omega}^\dag\),\qquad
y_\Omega=\frac{1}{i\sqrt2}\(\alpha_\Omega-\alpha_{-\Omega}^\dag\),\qquad x_{\Omega}=x^\ast_{-\Omega},\qquad
y_{\Omega}=y^\ast_{-\Omega}\L{37}
\EY
and $\alpha_\Omega$ is eigen-number of the annihilation operator $\hat A_\Omega$ (\ref{27}).

It is clear that this CM is quite different from discussed above (\ref{24}), although here and there we investigated the same object, namely the
system consisting of two field oscillators.

Note that in the paper \cite{6} we have studied theoretically triply resonant non-degenerate optical parametric oscillator operating above threshold
and constructed the CM in form (\ref{36}). In order to determine the quantum state purity, the formula (\ref{28}) was applied. Further we will show
that as far as this CM describes the two-oscillator system, the correct formula connecting it with the purity is different.

\subsection{Quantum state purity }

Since we consider two field oscillators with frequencies $ \omega_0 \pm \Omega $, then a purity of the quantum state in the Wigner representation can
be derived as an integral expression:
\BY
&&{\cal P}_{\Omega,-\Omega}=\pi^2\int\int d^2\alpha_\Omega
d^2\alpha_{-\Omega}\;W^2(\alpha_\Omega,\alpha_{-\Omega}),\L{33}
\EY
where $W(\alpha_\Omega,\alpha_{-\Omega})$ is the two mode Wigner quasi-distribution.

Taking into account that
\BY
&&\alpha_\Omega=\frac{1}{\sqrt2}\(x_\Omega+iy_\Omega\),
\EY
we can change the variables of integration in Eq. (\ref{33}) from $ \alpha_{\pm\Omega} $ to $ x_\Omega $ and $
y_{\Omega} $, then
\BY
&&{\cal P}_{\Omega,-\Omega}=\frac{\pi^2}{4}\int\int d^2x_\Omega d^2y_\Omega
\;W^2(\alpha_\Omega,\alpha_{-\Omega}).\L{40}
\EY
Assuming that the observed system of two oscillators obeys the Gaussian statistics, one  can derive the Wigner
quasi-distribution in general form:
\BY
&&W(\alpha_\Omega,\alpha_{-\Omega})=\frac{4}{\pi^2}(ab-c^2 )\exp\({ -a \;\delta x_\Omega\delta x_{-\Omega}-b \;\delta
y_\Omega\delta y_{-\Omega}+c \(\delta x_\Omega \delta y_{-\Omega}+\delta x_{-\Omega} \delta y_\Omega\) }\).\L{41}
\EY
Substituting this quasi-distribution into (\ref{40}) and carrying out an integration in the explicit form one can obtain ${\cal
P}_{\Omega,-\Omega}=ab-c^2$.

Calculating directly the momenta of the distribution (\ref{41}), one can get the following non-zero momenta
\BY
&&2\langle|\delta x_\Omega|^2\rangle=\frac{b}{ab-c^2} ,\qquad2\langle|\delta
y_\Omega|^2\rangle=\frac{a}{ab-c^2},\qquad 2\langle\delta x_\Omega\delta y_{-\Omega}\rangle=2\langle\delta
x_{-\Omega}\delta y_\Omega\rangle=\frac{c}{ab-c^2}.\L{}
\EY
Then the determinant of the CM (\ref{36}) equals to $\det{\cal M}_{\Omega,-\Omega}=1/(ab-c^2)$, and it is not difficult to see that
\BY
&&{\cal P}_{\Omega,-\Omega}= \frac{1}{\det {\cal M}_{\Omega,-\Omega}}.\L{35}
\EY
Unlike (\ref{28}) here there is no root in denominator.

As one can see, in order to reconstruct the Gaussian quasi-distribution (\ref{41}), we can exploit the CM (\ref{36}) and calculate the parameters
$a,\;b$ and $c$ according to equations
\BY
&&b=2\langle\delta x_\Omega\delta x_{-\Omega}\rangle/\det{\cal M}_{\Omega,-\Omega} ,\qquad a=2\langle\delta
y_\Omega\delta y_{-\Omega}\rangle/\det{\cal M}_{\Omega,-\Omega},\qquad c=2\langle\delta x_\Omega\delta
y_{-\Omega}\rangle/\det{\cal M}_{\Omega,-\Omega},\L{}
\EY
where
\BY
&&\det{\cal M}_{\Omega,-\Omega}=2\langle\delta x_\Omega\delta x_{-\Omega}\rangle2\langle\delta y_\Omega\delta
y_{-\Omega}\rangle- 2\langle\delta x_\Omega\delta y_{-\Omega}\rangle2\langle\delta x_{-\Omega}\delta y_\Omega\rangle,
\EY
In conclusion we would like to emphasize once more, as we have shown above the full reconstruction of the field state is achieved when we use the CM,
built on the real and imaginary parts of the spectral field amplitude. Although the investigation in terms of the quadrature components of the field
is very popular, nevertheless the corresponding CM does not give a possibility for the full reconstruction of the quantum state of the field. However
it can be productive for reconstruction of the two-oscillator systems with frequencies, placed symmetrically relative to mode frequency, because, in
this case, the corresponding CM is simpler and more convenient for calculation.

\section{ACKNOWLEDGEMENT\L{X}}

We thank Claude Fabre for productive discussion. The reported study was supported by RFBR (grants No. 12-02-00181a
and 13-02-00254a) and by the ERA.Net RUS Project NANOQUINT.

\end{document}